\documentstyle[12pt]{article}
\newcommand{\alt}{\mathbin{\lower 3pt\hbox
   {$\rlap{\raise 5pt\hbox{$\char'074$}}\mathchar"7218$}}}
\newcommand{\agt}{\mathbin{\lower 3pt\hbox
   {$\rlap{\raise 5pt\hbox{$\char'076$}}\mathchar"7218$}}}

\begin{document}
\setcounter{footnote}{0}
\setcounter{equation}{0}
\setcounter{figure}{0}
\setcounter{table}{0}
\vspace*{5mm}

\begin{center}
{\large\bf Renormalons and the Renormalization Scheme }

\vspace{4mm}
 I. M. Suslov \\
{ \it P.L.Kapitza Institute for Physical Problems,  119334 Moscow, Russia} \\
\vspace{6mm}

\begin{minipage}{135mm}
{\rm {\quad} The possibility is discussed that existence of renormalon
singularities is not the internal property of the specific field theory
but depends on the renormalization scheme.  }
\end{minipage} \end{center}

According to the recent paper \cite{1}, existence or absence of
renormalon singularities is related with the analiticity properties
of the Gell-Mann -- Low function $\beta(g)$  ($g$ is a coupling
constant). Briefly, the results are as follows:

(a) Renormalon singularities are absent, if $\beta(g)$ has a proper
behavior at infinity, $\beta(g)\sim g^\alpha$ with $\alpha\le 1$,
and its singularities at finite points $g_c$ are sufficiently weak,
so that $1/\beta(g)$ is not integrable at $g_c$
(i.e. $\beta(g)\sim (g-g_c)^\gamma$ with $\gamma>1$).

(b) Renormalon singularities exist, if at least one condition
named in (a) is violated.

It is well-known \cite{2} that the Gell-Mann -- Low function
$\beta(g)$ depends on the renormalization scheme, and only two
coefficients $\beta_2$ and $\beta_3$ are universal in the expansion
$\beta(g)=\beta_2 g^2+ \beta_3 g^3+ \ldots$ In essence, the
change of the renormalization scheme is simply a change of variables
$g=f(\tilde g)$, transferring $\beta(g)$ to
$\tilde \beta(\tilde g)=\beta(f(\tilde g))/f'(\tilde g) $.
Function  $f(g)$ is subjected to certain
physical restrictions, such as $f(g)=g +O(g^2)$; in fact, these restrictions
are poorly investigated.  The interesting possibility arises, if these
restrictions do not forbid to transform the $\beta$ function of type (a) to the
$\beta$ function of type (b). In this case, existence or absence of
renormalon singularities is not {\it the internal property} of the specific
field theory but depends on the renormalization scheme, i.e. on {\it the
way of description}.\footnote{\,An analogous conclusion was drawn in Ref.3
in result of more tedious and less rigorous analysis.}
The observable quantities do not depend on the
renormalization scheme and the latter can be chosen from  convenience.

On the one hand, the scheme without renormalons can be used to formulate
the well-defined theory with unique predictions \cite{1,3}. In such a
theory, large orders of  perturbation expansion are determined by the
Lipatov method, the Borel integral is well-defined and constructive
summation of the perturbation series is possible, giving the possibility
to solve different strong coupling problems \cite{3}. It was argued
in \cite{1,3} that the $MOM$ scheme in  $\phi^4$ theory and the
$MS$ scheme in QED and QCD are renormalon-free.

On the other hand, one can deliberately choose an "extremely renormalon"
scheme, in order to justify the renormalon heuristics, which is extensively
used in  different applications \cite{4}. For example,  power corrections
in QCD are determined generally by the wide set of diagrams and can be
calculated starting from the "renormalon end" \cite{4} or from the "instanton
end" \cite{5}. When the $\beta$ function of type (a) is used, the main
contribution to  power corrections is determined by instantons;  when
the $\beta$ function is of type (b), this contribution is localized on the
"renormalon end" and can be argued to possess "universality"
\cite{6}, etc.  It is interesting, that experiment seems to agree with the
instanton models, as well as with "renormalon universality" \cite{7}.

Author is indebted to A.L.Kataev for stimulating discussions.


\end{document}